\documentclass[conference,a4paper]{IEEEtran}
\IEEEoverridecommandlockouts

\usepackage{cite}
\usepackage{amsmath,amssymb,amsfonts}
\usepackage{algorithmic}
\usepackage{graphicx}
\usepackage{textcomp}
\usepackage{xcolor}
\usepackage{verbatim}
\usepackage{booktabs}
\usepackage[table]{xcolor}
\def\BibTeX{{\rm B\kern-.05em{\sc i\kern-.025em b}\kern-.08em
    T\kern-.1667em\lower.7ex\hbox{E}\kern-.125emX}}
\begin{document}

\makeatletter
\def\ps@IEEEtitlepagestyle{%
  \def\@oddfoot{\mycopyrightnotice}%
  \def\@evenfoot{}%
}
\def\mycopyrightnotice{%
  {\footnotesize   \hfill} 
}
\makeatother

\title{Interference-Driven Clustered Optimisation
for FM Spectrum Coordination\\
\thanks{This work has been carried out in the framework of a project between the Ministry of "Imprese e Made in Italy" and Fondazione Ugo Bordoni.}
}

\author{\IEEEauthorblockN{Federica Mangiatordi, Emiliano Pallotti}
\IEEEauthorblockA{\textit{Fondazione Ugo Bordoni } \\
Viale del Policlinico 147, Rome, Italy \\
\{fmangiatordi,epallotti\}@fub.it}
}

\maketitle
\begin{abstract}

Cross-border FM spectrum coordination involves protecting foreign broadcasting services while preserving domestic coverage, amid increasingly large radio-planning datasets containing thousands of transmitters and millions of transmitter-pixel relationships. In such scenarios, conventional optimisation approaches become computationally demanding due to the high dimensionality of the associated power-control problem.
This paper proposes an interference-driven clustered optimisation framework for large-scale FM spectrum coordination. The proposed method exploits the observation that violations of foreign-service protection are typically dominated by a limited subset of transmitters. Protected services are therefore analysed to identify dominant interferers and quantify their impact on interference. These relationships are represented through an interference graph from which optimisation-oriented transmitter clusters are extracted.
The clusters decompose the global power-control problem into smaller optimisation tasks solved with clustered simulated annealing, followed by a global refinement that captures residual inter-cluster interactions. Coverage and interference are evaluated using frequency-dependent protection criteria and a dynamic strongest-service assignment model.
To enable operational-scale planning, the framework uses sparse matrices and GPU-accelerated computations. Tests on realistic cross-border FM coordination scenarios show that the clustering strategy greatly reduces optimisation complexity and runtime while maintaining foreign-service protection and domestic coverage. The method also yields an interpretable ranking of transmitters that contribute most to harmful interference, supporting optimisation and spectrum planning.

\end{abstract}


\section{Introduction}

Frequency Modulation (FM) broadcasting is a key component of national radio infrastructures and requires effective cross-border spectrum coordination to ensure compatibility between neighbouring administrations operating in shared frequency bands. Compliance with international protection requirements is evaluated with extensive radio-planning databases that store field-strength predictions for thousands of transmitters and millions of transmitter–pixel combinations \cite{BS412,Philipp2011,GE84,mappatao2010radiation}. Consequently, operational FM coordination involves large-scale optimisation problems whose computational complexity rapidly increases with the size of the planning dataset.

In practical coordination, cross-border protection violations usually stem from a small set of dominant interferers, so optimisation can focus on the transmitters mainly responsible for harmful interference rather than the entire network.
Several optimisation techniques have been proposed for radio planning, including mathematical programming, evolutionary algorithms and metaheuristics \cite{10431770,PPR949811,ChavesGonzalez2010,Aardal2003FAP}. More recently, graph-based representations have been investigated for analysing interference relationships and decomposing large optimisation problems into smaller, more tractable subproblems \cite{Wang2022RCNetDecomp,Dai2025GraphSurvey}.
Nevertheless, existing approaches do not explicitly exploit the sparse, violation-driven interference structure of operational cross-border FM coordination, motivating the proposed interference-driven decomposition.

This paper proposes an interference-driven clustered optimisation framework for large-scale FM spectrum coordination. Dominant interferers are first identified from the observed protection violations and represented through an interference graph. The resulting graph is decomposed into optimisation-oriented transmitter clusters that are processed sequentially using Simulated Annealing, followed by a final refinement stage. The proposed framework combines graph decomposition, sparse matrix representations and GPU acceleration \cite{11268788,9626937} to efficiently handle operational radio-planning datasets.
The framework specifically targets operational cross-border FM spectrum planning rather than cellular network architectures.
The main contributions of this work are:

\begin{itemize}

\item an interference-impact ranking methodology for identifying the dominant transmitters responsible for foreign-service protection violations;

\item an interference-driven graph decomposition strategy that constructs optimisation-oriented transmitter clusters;

\item a clustered optimisation framework based on Simulated Annealing for large-scale FM spectrum coordination.

\end{itemize}
The remainder of this paper is organised as follows. Section~II describes the system model adopted for FM spectrum coordination. Section~III presents the proposed interference-driven clustered optimisation framework. Section~IV discusses the experimental results obtained on a realistic operational coordination scenario. Finally, Section~V concludes the paper.

\section{System Model and Problem Formulation}

\subsection{Network Scenario}

We consider a cross-border FM broadcasting scenario with two neighbouring administrations, referred to as Country~A and Country~B. The objective is to coordinate controllable transmitters in Country~A  to reduce harmful interference to radio services in Country~B while preserving domestic coverage.
The geographical area is discretised into pixels from a radio-planning database. Let $\mathcal{P}^{A}$ and $\mathcal{P}^{B}$ be the pixel sets of Country~A and Country~B, respectively. The controllable and protected transmitters are denoted by $\mathcal{T}^{A}$ and $\mathcal{T}^{B}$, respectively, with the operating parameters of $\mathcal{T}^{B}$ fixed during optimisation.
For each transmitter $j$ and pixel $p$, the radio-planning database provides 
\begin{equation}
E_{j,p}^{(50)},
\end{equation}
in dB$\mu$V/m 
representing the median predicted field strength at the considered location, in accordance with the FM planning field-strength statistics of ITU-R BS.412 \cite{BS412}.
Power control is represented by
\begin{equation}
\mathbf{x}=
[x_1,x_2,\ldots,x_{N_T}]^{T},
\end{equation}
where $x_j\le 0$ is the power reduction  applied to transmitter $j\in\mathcal{T}^{A}$. 
Define
\begin{equation}
y_j = 10^{x_j/10},
\qquad
0<y_j\le1,
\end{equation}
so the corresponding linearised field-strength contribution is
\begin{equation}
w_{j,p}(\mathbf{x})
=
10^{E_{j,p}^{(50)}/10}\cdot y_j .
\label{eq:field_linear}
\end{equation}
Here $w_{j,p}(\mathbf{x})$ is proportional to the received-power contribution from
transmitter $j$ at pixel $p$, allowing useful and interfering contributions to be summed linearly. Only transmitters in Country~A are optimisation variables.

\subsection{Foreign Service Protection Model}

Each transmitter in Country~B is treated as an independent protected service and evaluated over all pixels where its field strength exceeds the minimum service threshold.

For a protected transmitter $m\in\mathcal{T}^{B}$ and pixel $p$, the useful contribution is
\begin{equation}
C_{m,p}=
10^{E_{m,p}^{(50)}/10}, \quad  m\in\mathcal{T}^{B}
\end{equation}
Interference is evaluated by considering co-channel and adjacent-channel contributions. Let
\begin{equation}
\mathcal{K}=\{0,100,200\}
\end{equation}
denote the set of frequency separations (kHz), corresponding to co-channel, first-adjacent ($\pm100$ kHz), and second-adjacent ($\pm200$ kHz) relationships. For each $k\in\mathcal{K}$, let $\mathcal{I}_k(m,p)$ denote the set of controllable transmitters whose carrier frequency differs from that of transmitter $m$ by $k$ kHz.
The aggregate interference is
\begin{equation}
I_{m,p}(\mathbf{x})
=
\sum_{k\in\mathcal{K}}
\alpha_k
I_{m,p}^{(k)}(\mathbf{x}),
\label{eq:total_interference}
\end{equation}
with
\begin{equation}
I_{m,p}^{(k)}(\mathbf{x})
=
\sum_{j\in\mathcal{I}_k(m,p)}
w_{j,p}(\mathbf{x}),
\qquad
k\in\mathcal{K},
\label{eq:band_interference}
\end{equation}
and
\begin{equation}
\label{eq:alphak}
\alpha_k = 10^{M_k/10},
\end{equation}
where $M_k$ is the protection ratio associated with frequency separation $k$.
A service pixel is considered protected if
\begin{equation}
C_{m,p}
\ge
\Gamma_{\mathrm{EXT}}
I_{m,p}(\mathbf{x}),
\label{eq:protection_constraint}
\end{equation}
where $\Gamma_{\mathrm{EXT}}=1$ (0 dB), since the frequency-dependent protection requirements are already incorporated into $\alpha_k$.
Residual violations are quantified through
\begin{equation}
\xi_{m,p}
=
\max
\left(
\Gamma_{\mathrm{EXT}}
I_{m,p}(\mathbf{x})
-
C_{m,p},
0
\right).
\label{eq:slack}
\end{equation}
A value $\xi_{m,p}=0$ means the protection criterion is met, while positive values indicate the violation severity. The overall protection performance of a candidate power-control solution is obtained by aggregating pixel-wise violations over all protected transmitters and service pixels.

\subsection{National FM Service Evaluation}

The national FM service is evaluated using the same interference model adopted for foreign-service protection, combined with a dynamic best-server assignment.
For each pixel $p\in\mathcal{P}^{A}$ and frequency $f$, let

\begin{equation}
\mathcal{T}^{A}(p,f)
=
\left\{
j :
f_j=f
\ \text{and} \
w_{j,p}(\mathbf{x})>0
\right\},
\label{eq:it_tx_set}
\end{equation}

denote the set of Country~A transmitters received at pixel $p$ on frequency $f$. The serving transmitter is selected as

\begin{equation}
b(p,f)
=
\arg\max_{j\in\mathcal{T}^{A}(p,f)}
w_{j,p}(\mathbf{x}).
\label{eq:best_server}
\end{equation}.

The corresponding useful signal is

\begin{equation}
C_{p,f}(\mathbf{x})
=
w_{b(p,f),p}(\mathbf{x}).
\label{eq:useful_signal_it}
\end{equation}

The aggregate interference is computed according to the same co-channel and adjacent-channel model introduced in Section II-B,

\begin{equation}
I_{p,f}(\mathbf{x})
=
\sum_{k\in\mathcal{K}}
\alpha_k\,I^{(k)}_{p,f}(\mathbf{x}),
\label{eq:interference_it}
\end{equation}

with

\begin{equation}
I^{(k)}_{p,f}(\mathbf{x})
=
\sum_{j\in\mathcal{I}_k(p,f)}
w_{j,p}(\mathbf{x}),
\qquad
k\in\mathcal{K}.
\label{eq:interference_it_components}
\end{equation}

A pixel--frequency pair is considered covered when both

\begin{equation}
C_{p,f}(\mathbf{x})
\ge
\Gamma_A I_{p,f}(\mathbf{x}),
\label{eq:ci_condition}
\end{equation}

and

\begin{equation}
E^{(50)}_{b(p,f),p}
\ge
E_{\min}^{A},
\label{eq:field_strength_condition}
\end{equation}

are satisfied, where $\Gamma_A=1$ (0 dB) and $E_{\min}^{A}$ denotes the minimum service field strength specified by the international FM planning criteria\cite{BS412,Philipp2011}.
Residual violations are quantified through the slack variable

\begin{equation}
z_{p,f}(\mathbf{x})
=
\max
\left(
\Gamma_A I_{p,f}(\mathbf{x})
-
C_{p,f}(\mathbf{x}),
0
\right),
\label{eq:slack_it}
\end{equation}
which is zero when the reception criterion is satisfied and otherwise measures the severity of the violation.

\subsection{Sparse Interference Representation}

The interference relationships between controllable transmitters in Country~A and protected services in Country~B are represented through sparse matrices derived from the radio-planning database. Each protected transmitter--pixel pair corresponds to one matrix row, whereas each controllable transmitter corresponds to one column.

Three sparse matrices,
$A_0$, $A_{100}$ and $A_{200}$,
represent the co-channel, first-adjacent ($\pm100$ kHz) and second-adjacent ($\pm200$ kHz) interference contributions, respectively.

Let $\mathbf{y}$ denote the vector collecting the linear power-scaling factors introduced in (\ref{eq:field_linear}). The aggregate interference affecting all protected transmitter--pixel pairs can then be written in compact form as

\begin{equation}
\mathbf{I}
=
\alpha_0A_0\mathbf{y}
+
\alpha_{100}A_{100}\mathbf{y}
+
\alpha_{200}A_{200}\mathbf{y}
+
\mathbf{I}_{\mathrm{fixed}},
\label{eq:sparse_interference}
\end{equation}

where $\mathbf{I}_{\mathrm{fixed}}$ represents the aggregate contribution of non-controllable transmitters.

In addition to facilitating efficient sparse matrix–vector computations on graphics processing units (GPUs), the non-zero coefficients of $A_0$, $A_{100}$, and $A_{200}$ inherently induce a bipartite graph that connects controllable transmitters to protected transmitter–pixel pairs. This graph serves as the foundation for the interference-driven clustering methodology presented in the subsequent section.

\section{Proposed Interference-Driven Clustered Optimisation}

\begin{figure}
    \centering
    \includegraphics[width=1\linewidth]{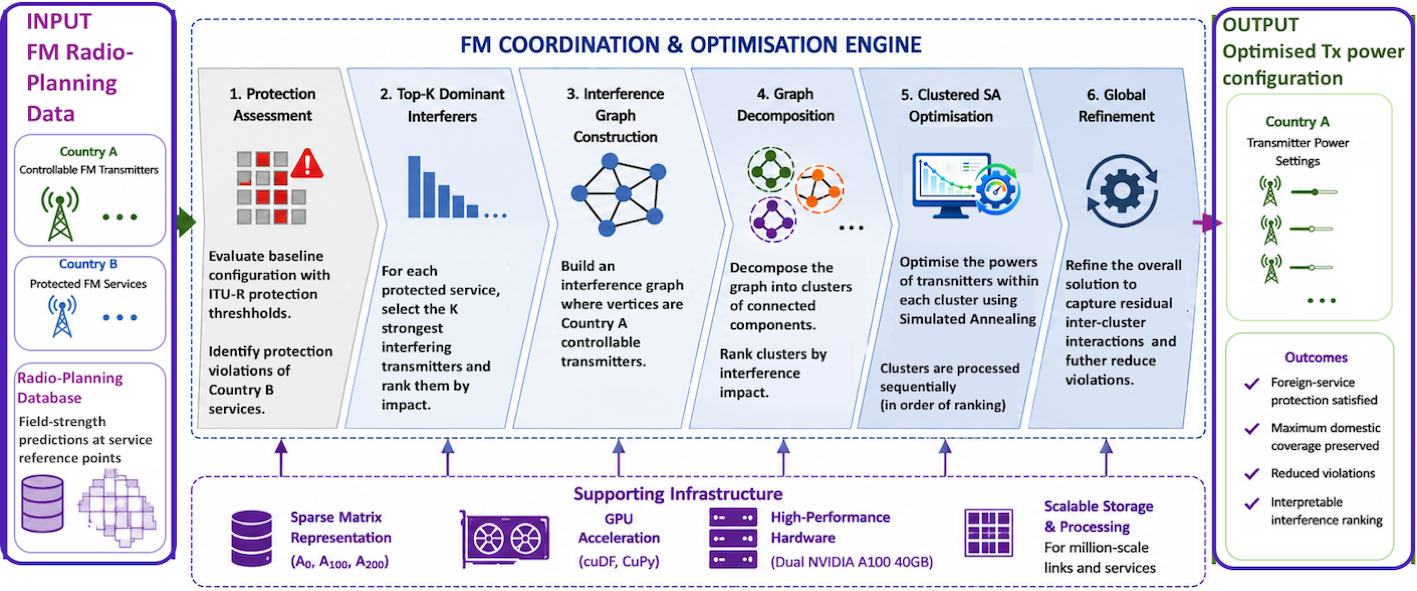}
    \caption{Architecture of the proposed FM coordination framework}
    \label{fig:workflow}
\end{figure}

The proposed framework exploits the sparse structure of operational FM interference, where cross-border protection violations are typically dominated by a limited number of interfering transmitters. Instead of optimising all controllable transmitters simultaneously, the method identifies the dominant interferers, represents their relationships through an interference graph, and decomposes the original coordination problem into a sequence of smaller optimisation tasks. The overall workflow is illustrated in Fig.~\ref{fig:workflow}.
\subsection{Identification and Ranking of Dominant Interferers}

The process starts from the nominal operating configuration. Foreign-service protection is evaluated according to the violation metric defined in (\ref{eq:slack}), and only protected transmitter--pixel pairs satisfying
\begin{equation}
\xi_{m,p}>0
\end{equation}
are retained for further analysis.

For each protected transmitter $m$, the individual interference contributions generated by controllable transmitters in Country~A are ranked. The $K$ strongest contributors define the set of dominant interferers:
\begin{equation}
\mathcal{I}_m^{\mathrm{Top}K}
=
\{i_1,i_2,\ldots,i_K\}.
\end{equation}

To quantify the relevance of each controllable transmitter $i$ across all protected services, an occurrence score is introduced:
\begin{equation}
R_i
=
\sum_{m\in\mathcal{T}^{B}}
\mathbf{1}
\left(
i\in\mathcal{I}_m^{\mathrm{Top}K}
\right),
\label{eq:ranking_score}
\end{equation}
where $\mathbf{1}(\cdot)$ is the indicator function. A larger value of $R_i$ indicates that transmitter $i$ appears more frequently among the dominant interferers and is therefore expected to have a stronger impact on the coordination process.

\subsection{Interference Graph Construction}

The dominant-interferer relationships are represented through an undirected graph whose vertices correspond to controllable transmitters in Country~A. Two transmitters are connected when they jointly appear among the Top-$K$ interferers of at least one protected service in Country B:
\begin{equation}
\exists\,m:
\;
i\in\mathcal{I}_m^{\mathrm{Top}K}
\land
j\in\mathcal{I}_m^{\mathrm{Top}K}.
\end{equation}

The connected components of this graph define the optimisation clusters:
\begin{equation}
\mathcal{C}_1,\mathcal{C}_2,\ldots,\mathcal{C}_N .
\end{equation}

Each cluster groups transmitters that jointly affect common protected services. To prioritise the optimisation sequence, cluster $\mathcal{C}_k$ is assigned the score
\begin{equation}
S_k
=
\sum_{i\in\mathcal{C}_k}
R_i,
\end{equation}
and clusters are processed in decreasing order of $S_k$.

\subsection{Optimisation Problem}

Let $\mathbf{x}$ denote the vector of power reductions applied to the controllable transmitters of Country~A, and let $\mathbf{y}$ be the corresponding vector of linear scaling factors, with $y_j=10^{x_j/10}$.

The optimisation objective combines three terms:
\begin{equation}
J(\mathbf{x})=
\lambda_B\Phi_B(\mathbf{x})
+
\lambda_A\Phi_A(\mathbf{x})
+
\lambda_P\Phi_P(\mathbf{x}),
\label{eq:objective_function}
\end{equation}
where $\Phi_B$ penalises residual foreign-service protection violations, $\Phi_A$ penalises degradation of national FM service availability, and $\Phi_P$ limits unnecessary power reductions.

The foreign-service term is defined as
\begin{equation}
\Phi_{B}(\mathbf{x})=
\sum_{m\in\mathcal{T}^{B}}
\sum_{p\in\mathcal{P}^{B}_{m}}
\xi_{m,p}(\mathbf{x}),
\end{equation}
while the national-service term is
\begin{equation}
\Phi_{A}(\mathbf{x})=
\sum_{(p,f)\in\mathcal{Q}^{A}}
z_{p,f}(\mathbf{x}).
\end{equation}
Finally, the power-reduction penalty is
\begin{equation}
\Phi_{P}(\mathbf{x})=
\sum_{j\in\mathcal{T}^{A}}
\left(1-y_j\right).
\end{equation}

The weighting coefficients $\lambda_B$, $\lambda_A$, and $\lambda_P$ control the relative importance of foreign-service protection, national service preservation, and limitation of power reductions.

\subsection{Clustered Simulated Annealing}

Each optimisation cluster is processed using a Simulated Annealing procedure \cite {kirkpatrick1983sa} that minimises the objective function in (\ref{eq:objective_function}). During the optimisation of cluster $\mathcal{C}_k$, only the transmitters belonging to that cluster are allowed to vary, while all remaining transmitters remain fixed. The evaluation domain, however, remains the complete coordination scenario, including all protected services and national coverage constraints.

Candidate solutions are generated by randomly perturbing the power reductions of the active transmitters. Worsening moves are accepted according to the Metropolis criterion \cite{kirkpatrick1983sa}:
\begin{equation}
P_{\mathrm{acc}}
=
\exp
\left(
-\frac{\Delta J}{T}
\right),
\qquad
\Delta J>0.
\end{equation}

The optimised solution obtained for cluster $\mathcal{C}_k$ is used as the initial condition for the optimisation of cluster $\mathcal{C}_{k+1}$. This sequential strategy progressively reduces harmful interference while keeping the dimensionality of each optimisation subproblem substantially smaller than that of the monolithic formulation.

\subsection{Global Refinement}

The Top-$K$ graph intentionally focuses the optimisation on the dominant interference relationships. However, weaker non-clustered transmitters may still contribute to the overall interference level.
For this reason, a final refinement stage is performed after the clustered optimisation. The same objective function and Simulated Annealing procedure are used, but the optimisation variables are restricted to the controllable transmitters not included in any cluster. The power reductions previously obtained for clustered transmitters remain fixed.
This final stage compensates for the approximation introduced by the Top-$K$ representation and accounts for residual interference contributions without increasing the dimensionality of the cluster-level optimisation problems.

\section{Experimental Results}

The proposed methodology was evaluated in a real-world operational FM spectrum coordination scenario to assess its effectiveness in enhancing foreign-service protection while preserving the availability of the national FM network.

\subsection{Experimental Scenario}

The proposed methodology was evaluated on a real operational cross-border FM coordination scenario derived from radio-planning databases. For confidentiality reasons, the neighbouring administrations are denoted as Country~A and Country~B.
The analysed area covers approximately $15\,766\,\mathrm{km}^{2}$ with a spatial resolution of about $2\,\mathrm{km}^{2}$. The optimisation simultaneously considers 3\,247 controllable transmitters in Country~A and 630 protected FM services in Country~B, resulting in nearly 94\,000 candidate service points and more than seven million co-channel and adjacent-channel interference relationships.
Coverage and compatibility analyses were performed in accordance with ITU-R Recommendation BS.412. The main characteristics of the analysed coordination scenario are summarised in Table~\ref{tab:scenario}.
\begin{table}[h]
\centering
\scriptsize
\setlength{\tabcolsep}{4pt}
\caption{Characteristics of the analysed coordination scenario.}
\label{tab:scenario}
\footnotesize
\setlength{\tabcolsep}{4pt}
\begin{tabular}{lr|lr}
\hline
Metric & Value & Metric & Value\\
\hline
Study area & 15\,766 km$^2$ &
Pixel size & 2 km$^2$\\

Country~A pixels & 4\,576 &
Country~B pixels & 3\,307\\

Controllable transmitters & 3\,247 &
Protected FM services & 630\\

Candidate service points & 93\,977 &
Co-channel links & 1\,389\,679\\

$\pm100$ kHz links &
2\,856\,299 &
$\pm200$ kHz links &
2\,837\,655\\
\hline
\end{tabular}
\end{table}
\subsection{Maximum Recoverable Foreign-Service Area}

Before comparing different optimisation strategies, the theoretical upper bound achievable through transmitter power control must first be established.

The protected service area in Country~B is not determined solely by cross-border interference generated by transmitters in Country~A. It is also constrained by the characteristics of the protected FM network itself, including transmitter deployment, frequency assignments and internal interference. Consequently, a fraction of the service area remains unavailable even if all controllable transmitters in Country~A are switched off.
To distinguish recoverable from non-recoverable service losses, a reference scenario was computed by completely disabling all controllable transmitters in Country~A. Since no further improvement in transmitter power control can be achieved beyond this condition, the resulting protected service area defines the theoretical upper bound of the optimisation problem. Throughout the remainder of this paper, this reference is referred to as the \emph{maximum recoverable foreign-service area} and is used as the benchmark for evaluating the effectiveness of the proposed optimisation framework.

\subsection{Foreign-Service Protection Performance}

Table~\ref{tab:coverages} compares the protected foreign-service area obtained under the investigated coordination strategies.

The baseline planning configuration satisfies the ITU-R BS.412 protection criterion for 16\,093 protected service pixels, corresponding to approximately 70.7\% of the maximum recoverable foreign-service area established in the previous subsection.

The proposed Clustered Simulated Annealing increases the protected foreign-service area to 22\,691 protected service pixels, corresponding to 99.68\% of the theoretical upper bound. The Full-Scale Simulated Annealing (Full Scale-SA), which optimises the same objective function over the complete set of more than 3\,200 controllable transmitters, protects 22\,698 service pixels, only seven more than the clustered approach. This corresponds to less than 0.03\% of the theoretical upper bound and is therefore operationally negligible for the analysed coordination scenario.
These results indicate that cross-border protection violations are largely governed by a limited subset of dominant interferers. Once these transmitters are represented in the interference graph, nearly the entire recoverable foreign-service area can be restored without simultaneously optimising all controllable transmitters. The close agreement between the clustered and monolithic solutions confirms that the proposed graph representation preserves the dominant interference mechanisms governing compliance with the BS.412 protection criterion.

\begin{table}[h]
\centering
\caption{Comparison of the investigated coordination strategies in terms of protected foreign-service area.}
\label{tab:coverages}
\footnotesize
\setlength{\tabcolsep}{2pt}
\begin{tabular}{lccc}
\hline
Configuration &
Protected service &
Recovery &
Gain vs. \\
&
pixels &
(\%) &
baseline \\
\hline
Baseline &
16\,093 &
70.70 &
-- \\

\rowcolor{gray!15}
Clustered SA ($K=30$) &
22\,691 &
\textbf{99.68} &
41.0\% \\

Full Scale-SA &
22\,698 &
99.71 &
41.2\% \\

\rowcolor{gray!15}
Upper Bound 
(Country~A TX Off) &
22\,763 &
100.00 &
41.4\% \\
\hline
\end{tabular}
\end{table}

\subsection{Influence of the Top-$K$ Parameter}

The Top-$K$ parameter sets how many dominant interferers are retained per protected service, controlling the detail of the cross-border interference model.
Table~\ref{tab:k_analysis} summarises the resulting interference graphs. As $K$ increases, more interferers are included, connected components grow, and more controllable transmitters enter the optimisation. The graph thus evolves from many small, nearly independent clusters towards a structure resembling the full coordination problem.

Table~\ref{tab:k_results} reports the corresponding coordination results. The protected foreign-service area grows rapidly as $K$ increases, showing that the dominant interference mechanisms are progressively captured. Beyond about $K=20$, the recovered service area clearly saturates, while the optimisation problem keeps growing as weaker interferers are added.

Among the tested values, $K=30$ gives the best balance between performance and computational effort. It protects 22\,691 service pixels—99.68\% of the maximum recoverable foreign-service area—only seven pixels fewer than the Full-Scale Simulated Annealing solution.

The slight performance drop for $K>30$ stems from the increasing dimensionality of the optimisation problem: once dominant interference relationships are included, adding weaker interferers expands the search space without a commensurate increase in recoverable service area, thereby reducing the effectiveness of stochastic optimisation within a fixed computational budget.

These results show that compliance with the BS.412 protection criterion is mainly driven by a relatively small subset of dominant interfering transmitters; the rest contribute comparatively little. This supports the interference-driven graph decomposition and explains why the clustered optimisation can closely match the monolithic Full-Scale optimisation while solving much smaller problems.
\begin{table}[h]
\centering
\caption{Influence of the Top-$K$ parameter on the interference graph.}
\label{tab:k_analysis}
\scriptsize
\setlength{\tabcolsep}{2pt}
\renewcommand{\arraystretch}{0.82}
\begin{tabular}{cccc}
\hline
Top-$K$ &
Clusters &
Included
transmitters  &
Max. cluster size \\
&
&
(\%) &
(TX) \\
\hline
5  & 97 & 27.30 & 24   \\
10 & 28 & 44.34 & 144  \\
15 &  8 & 54.85 & 537  \\
20 &  3 & 61.65 & 1087 \\
25 &  3 & 65.84 & 1308 \\
30 &  2 & 69.23 & 1807 \\
35 &  1 & 71.88 & 2334 \\
40 &  1 & 73.85 & 2398 \\
\hline
\end{tabular}
\end{table}

\begin{table}[t]
\centering
\caption{Influence of Top-$K$ on coordination performance.}
\label{tab:k_results}
\scriptsize
\setlength{\tabcolsep}{2pt}
\renewcommand{\arraystretch}{0.82}
\begin{tabular}{cccc}
\hline
$K$ & Protected-service & Gap to upper bound(pixels) & Runtime (s) \\
\hline
5  & 22\,593 & -170 & 290  \\
10 & 22\,648 & -115 & 1665 \\
15 & 22\,678 & -85  & 717  \\
20 & 22\,690 & -73  & 406  \\
25 & 22\,690 & -73  & 475  \\
\textbf{30} & \textbf{22\,691} & \textbf{-72} & \textbf{324} \\
35 & 22\,684 & -79 & 210  \\
40 & 22\,674 & -89 & 194  \\
\textbf{Full Scale-SA} & \textbf{22\,698} & \textbf{-65} & \textbf{5517} \\
Upper bound & 22\,763 & 0 & -- \\
\hline
\end{tabular}
\end{table}

\subsection{Impact on National FM Service Availability}

Although the optimisation primarily targets foreign-service
protection, the resulting changes in transmitter power also alter interference within the national FM network.

To assess this, the national FM network was re-evaluated after optimisation using the same reception criterion as in the baseline scenario. Instead of analysing individual transmitters, the assessment counts the total number of covered pixel–frequency combinations across all national FM services. Because multiple services on different frequencies may provide reception at the same location, this metric reflects overall service availability rather than geographical coverage.
The results in Table~\ref{tab:domestic_availability} show that all optimisation strategies substantially increase aggregate service availability relative to the baseline. For the selected operating point ($K=30$), Clustered Simulated Annealing raises the number of covered pixel–frequency combinations from 10\,315 to 17\,418, a gain of about 69\%. Full-Scale Simulated Annealing achieves a similar outcome with only a small additional increase.

This gain does not expand the geographical coverage area; instead, it reduces co-channel and adjacent-channel interference in the national FM network, enabling more FM services to meet the reception criterion at the same locations. The optimisation thus reorganises the strongest-service assignment among transmitters rather than extending coverage.

The close match between clustered and Full-Scale optimisation confirms that the interference-driven graph decomposition preserves both foreign-service protection and the overall behaviour of the national FM network.

\begin{table}[t]
\centering
\caption{Aggregate national FM service availability for the investigated coordination strategies.}
\label{tab:domestic_availability}
\scriptsize
\setlength{\tabcolsep}{2pt}
\renewcommand{\arraystretch}{0.85}
\begin{tabular}{lcc}
\hline
Configuration & Covered pairs & Increase vs. baseline \\
\hline
Baseline Planning      & 10\,315 & -- \\
Clustered SA ($K=20$)   & 17\,554 & +70.2\% \\
Clustered SA ($K=25$)   & 17\,484 & +69.5\% \\
Clustered SA ($K=30$) & 17\,418 & +68.9\% \\
Full Scale-SA        & 17\,970 & +74.2\% \\
\hline
\end{tabular}
\end{table}
\begin{figure}[h]
    \centering
    \includegraphics[width=0.77\linewidth]{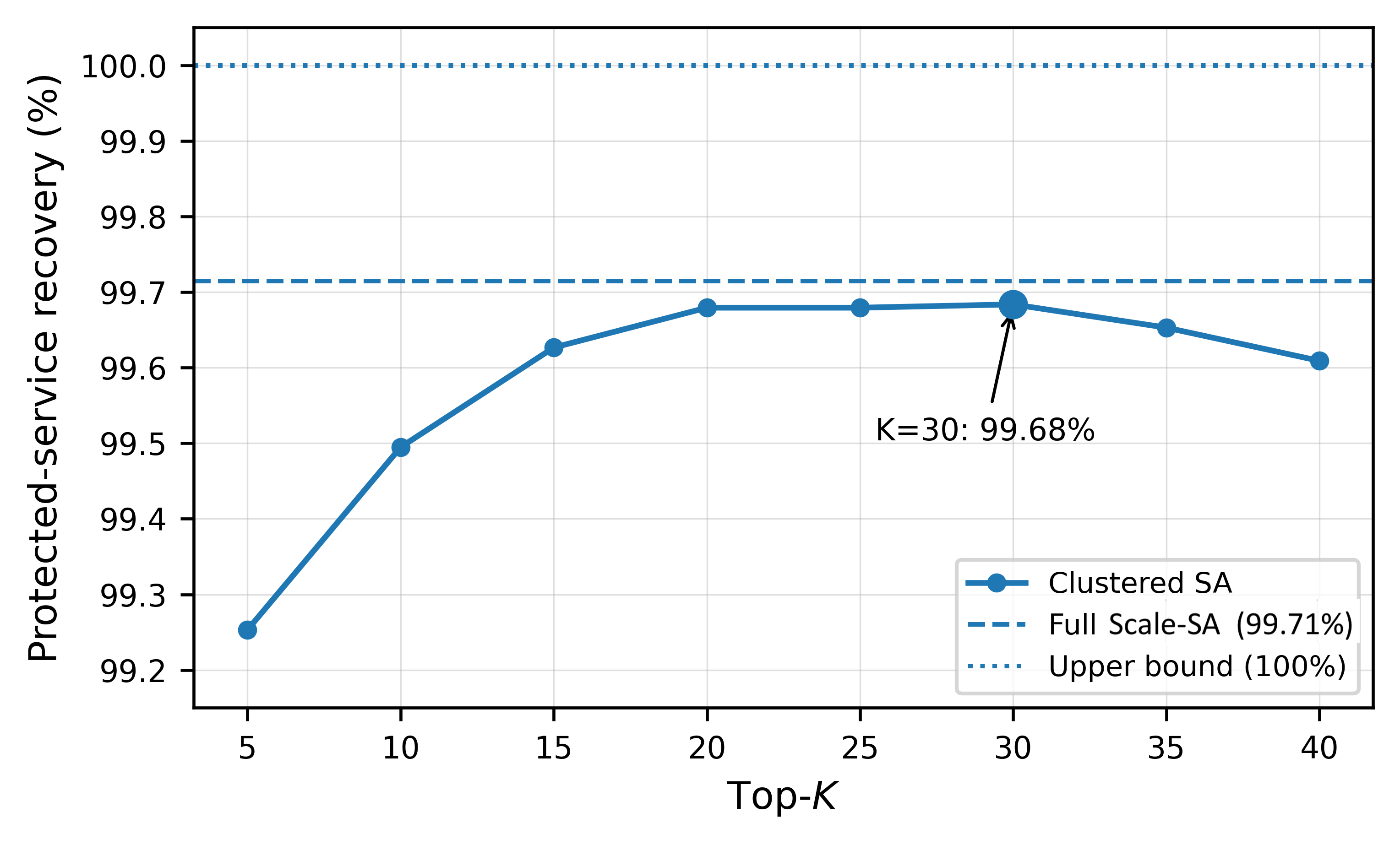}
    \caption{Protected foreign-service recovery versus Top-$K$.}
    \label{fig:placeholder}
\end{figure}
\subsection{Computational Performance}

Table~\ref{tab:k_results} reports the execution time of the proposed Clustered Simulated Annealing for the investigated values of the Top-$K$ parameter. Although the computational cost varies with the size of the interference graph induced by $K$, all clustered configurations require only a small fraction of the execution time of the Full-Scale Simulated Annealing (Full Scale-SA).
Among the investigated configurations, $K=30$ provides the best compromise between coordination performance and computational efficiency. At this operating point, the proposed clustered optimisation requires 324~s, compared with 5517~s for the Full Scale-SA, which is only 5.9\% of the monolithic optimisation's execution time.
The substantial reduction in computational cost is a direct consequence of the proposed interference-driven graph decomposition, which replaces a single large-scale optimisation problem involving all controllable transmitters with a sequence of significantly smaller optimisation problems. The experimental results demonstrate that this decomposition preserves nearly the same coordination performance while reducing computational time by approximately 94\%.

These results indicate that the proposed methodology provides a practical solution for large-scale operational FM frequency coordination, where computational efficiency is essential for analysing realistic planning scenarios.

\section{Conclusions}
The present paper has delineated an interference-driven graph decomposition framework for large-scale FM spectrum coordination. The proposed methodology exploits the observation that cross-border protection violations are primarily driven by a small set of dominant interferers. The identification and clustering of these transmitters enable the decomposition of the original coordination problem into a sequence of substantially smaller optimisation subproblems, while preserving the dominant interference mechanisms.
The proposed methodology was validated in a real operational FM coordination scenario involving several thousand controllable transmitters and millions of interference relationships. The clustered optimisation restored 99.68\% of the maximum recoverable foreign-service area. Concurrently, the national FM network was preserved through a reorganisation of the best-server assignment, rather than a degradation of the service provided.
From a computational perspective, the proposed graph decomposition dramatically reduces the optimisation effort. For the designated operating point ($K=30$), the clustered optimisation requires only 6\% of the runtime of the full-scale simulated annealing process, while achieving coordination performance indistinguishable from that of the latter.
The findings indicate that operational FM coordination challenges are predominantly driven by a small set of dominant interferers. The exploitation of this property enables near-optimal cross-border coordination while avoiding monolithic, large-scale optimisation, thereby making realistic coordination studies computationally practical.
\section*{Acknowledgement} 
The work has been carried out in the framework of the Spectrum Sharing project between the Ministry of Enterprises and Made in Italy (MIMIT) and Fondazione Ugo Bordoni.

\bibliographystyle{IEEEtran}
\bibliography{ References.bib}

@TechReport{GE84,
  author      = {{ITU}},
  title       = {Final Acts of the Regional Administrative Radio Conference for FM Sound Broadcasting (GE84)},
  institution = {ITU},
  year        = {1984}
}

@article{Philipp2011,
author = {Philipp, J.},
year = {2011},
month = {12},
pages = {391-396},
title = {Traditional protection ratios in FM sound broadcasting – still appropriate for interference management?},
volume = {9},
journal = {Advances in Radio Science},
doi = {10.5194/ars-9-391-2011}
}

@techreport{BS412,
  author       = {{International Telecommunication Union}},
  title        = {Planning Standards for Terrestrial FM Sound Broadcasting at VHF},
  institution  = {ITU-R},
  type         = {Recommendation ITU-R BS.412-9},
  address      = {Geneva, Switzerland},
  year         = {1998},
  month        = dec,
  note         = {Approved Dec. 14, 1998}
}

@ARTICLE{10431770,
  author={Avella, Pasquale and Nobili, Paolo and Sassano, Antonio},
  journal={IEEE Access}, 
  title={Power Reduction in FM Networks by Mixed-Integer Programming: A Case Study}, 
  year={2024},
  volume={12},
  number={},
  pages={23725-23732},
  doi={10.1109/ACCESS.2024.3365073}}

@inproceedings{mappatao2010radiation,
  title={Radiation pattern shaping for FM broadcast-optimising coverage},
  author={Mappatao, Gerino P},
  booktitle={2010 IEEE Symposium on Industrial Electronics and Applications (ISIEA)},
  pages={222--225},
  year={2010},
  organization={IEEE}
}

@misc {PPR949811,
	Title = {Holistic Interference Management for Wireless Networks in the Era of Artificial Intelligence},
	Author = {Husen, Arif and Nisar, Shafaq and Chaudary, Muhammad Hasanain and Khan, Zuhaib Ashfaq},
	DOI = {10.21203/rs.3.rs-5481165/v1},
	Publisher = {Research Square},
	Year = {2024},
}

@inproceedings{ChavesGonzalez2010,
  author    = {Chaves-Gonz{\'a}lez, J. M. and others},
  title     = {Swarm Intelligence, Scatter Search and Genetic Algorithm to Tackle a Realistic Frequency Assignment Problem},
  booktitle = {Distributed Computing and Artificial Intelligence},
  pages     = {441--448},
  year      = {2010}
}

@article{Aardal2003FAP,
  author  = {Karen Aardal and Stan van Hoesel and Arie M. C. A. Koster and others},
  title   = {Models and Solution Techniques for Frequency Assignment Problems},
  journal = {Annals of Operations Research},
  volume  = {153},
  number  = {1},
  pages   = {79--129},
  year    = {2007},
  doi     = {10.1007/s10479-007-0178-0}
}

@article{Wang2022RCNetDecomp,
  author  = {Junyuan Wang and Lin Dai and Lu Yang and Bo Bai},
  title   = {Clustered Cell-Free Networking: A Graph Partitioning Approach},
  journal = {IEEE Trans. Wireless Commun.},
  year    = {2023},
  doi     = {10.48550/arXiv.2207.11641}
}

@article{Dai2025GraphSurvey,
  author  = {Y. Dai and X. Shen},
  title   = {A Survey of Graph-Based Resource Management in Wireless Networks},
  journal = {IEEE Commun. Surveys Tuts.},
  year    = {2025}
}

@article{kirkpatrick1983sa,
  author  = {Kirkpatrick, S. and Gelatt, C. D. and Vecchi, M. P.},
  title   = {Optimization by Simulated Annealing},
  journal = {Science},
  volume  = {220},
  number  = {4598},
  pages   = {671--680},
  year    = {1983},
  doi     = {10.1126/science.220.4598.671}
}

@INPROCEEDINGS{9626937,
  author={Mangiatordi, Federica and Pallotti, Emiliano},
  booktitle={2021 AEIT International Annual Conference (AEIT)}, 
  title={A GPU accelerated framework for monitoring LTE/5G interference to DVB-T systems}, 
  year={2021},
  volume={},
  number={},
  pages={1-6},
  doi={10.23919/AEIT53387.2021.9626937}}

@INPROCEEDINGS{11268788,
  author={Pallotti, Emiliano and Faccioli, Manuel and Mangiatordi, Federica},
  booktitle={2025 17th International Congress on Ultra Modern Telecommunications and Control Systems and Workshops (ICUMT)}, 
  title={A High-Performance Framework for Large-Scale 5G/6G Terrestrial Coverage Simulation}, 
  year={2025},
  volume={},
  number={},
  pages={13-20},
  doi={10.1109/ICUMT67815.2025.11268788}}
\newpage
\end{document}